\documentclass[twocolumn,showpacs,preprintnumbers]{revtex4}
\usepackage{graphicx}
\usepackage{dcolumn}
\usepackage{bm}
\usepackage{epsfig}
\usepackage{amsmath}
\usepackage{subfigure}

\begin{document}

\title{Tunable single-photon multi-channel quantum router based on a
hybrid optomechanical system}
\author{Peng-Cheng Ma$^{1,2,3}$}
\author{Jian-Qi Zhang$^{2}$}
\email{boing777@qq.com}
\author{Yin Xiao$^1$}
\author{Mang Feng$^{2}$}
\email{mangfeng@wipm.ac.cn}
\author{Zhi-Ming Zhang$^{1}$}
\email{zmzhang@scnu.edu.cn}
\address{$^1$Laboratory of Nanophotonic Functional Materials and Devices (SIPSE), and
Laboratory of Quantum Engineering and Quantum Materials, South China
Normal University, Guangzhou 510006, China\\
$^2$State Key Laboratory of Magnetic Resonance and Atomic and Molecular Physics, Wuhan Institute of Physics and Mathematics,
Chinese Academy of Sciences, Wuhan 430071, China
\\ $^3$School of Physics
and Electronic Electrical Engineering, Huaiyin Normal University,
Huaian 223300, China}

\begin{abstract}
Routing$\wp_{l}$ of photon play a key role in optical communication
networks and quantum networks. Although the quantum routing of
signals has been investigated in various  systems both in theory and
experiment, the general form of quantum routing with multi-output
terminals still needs to be explored. Here, we propose an
experimentally accessible tunable single-photon multi-channel
routing scheme using a optomechanics cavity coulomb coupling to a
nanomechanical resonator. The router can extract a single-photon
from   the coherent input signal  directly modulate into three
different output channels.  More important, the two output signal
frequencies can be selected   by adjusting Coulomb coupling
strength. We also demonstrate the vacuum and thermal noise will be
insignificant for the optical performance of the single-photon
router at temperature of the order of 20 mK. Our proposal may have
paved a new avenue towards multi-channel router and quantum network.
\end{abstract}
 \maketitle

Quantum information science has been developed rapidly due to the
substitution of photons as signal carriers rather than the limited
electrons \cite{nature-453-1023}. Single photons are suitable
candidates as the carrier of quantum information due to the fact
that they propagate fast and interact rarely with the environment.
Meanwhile, a quantum single-photons router is challenging because
the interaction between individual photons is generally very weak.
Quantum router or quantum switch plays a key role in optical
communication  networks  and quantum information processing. It is
important for controlling the path of the quantum signal with fixed
Internet Protocol (IP) addresses, or quantum switch without fixed IP
addresses.



Designing a quantum router or  an optical switch operated at a
single photon level enables a selective quantum channel in quantum
information and quantum networks
\cite{prl-106-053901,prl-107-073601,pra-85-021801,nature-508-241,Science-341-768,
prl-111-193601, Nat.Photon-7-373, nat.phy.-3-807}, such as in
different systems of  quantum router, cavity QED system
\cite{prl-102-083601}, circuit QED system \cite{prl-107-073601},
optomechanical system \cite{pra-85-021801}, a pure linear optical
system \cite{pra-83-043814,arXiv:1207.7265}, $\wedge$-type
three-level system \cite{prx-3-031013,prl-111-103604,pra-89-013806}.
The essence lying at the core is the realization of the strong
coupling between the photons and photons or photons and phonons
\cite{prl-108-093604,Nat.Photon-6-605,Nat.Photon-4-477,Nat.Photon-2-185},
but these methods require high-pump-laser powers due to the very
weak optical nonlinearity. To the best of  our knowledge, the
quantum router demonstrated in most experiments and theoretical
proposals has only one output terminal, except for only a few
theoretical method in Ref.
\cite{prx-3-031013,prl-111-103604,pra-89-013806, scirep-4-4820} and
the experiments in Ref.\cite{arXiv:1207.7265}. However, the two
output ports are composited  the two photonic crystal cavities
\cite{prx-3-031013} or two  coupled-resonator waveguides
\cite{prl-111-103604,pra-89-013806} whose two output channels with a
maximal probability of unity and no more than $1/2$, respectively.
And the Ref.\cite{scirep-4-4820} is the extended to N output ports
from the method \cite{prl-111-103604}. Thus the ideal quantum router
with multi-channel deserves based on only one cavity with an
extremely high probability still need to be explored. In the
following, we propose a new scheme, via Coulomb coupling
interaction, one can realize one input signal with three output
signals based on only one optical cavity mode with unity
probability. Our proposal is essentially different from previous
methods in Ref.\cite{prx-3-031013,prl-111-103604,pra-89-013806} that
we are able to directly modulate the coherent input signal into
three different output ports.

   In this letter, we theoretical propose a scheme for single-photon
quantum routing with three output ports based on a optomechanics
cavity Coulomb coupling to a nanomechanical resonator (NMR). More
important, the two output signals frequencies can be selected  by
adjusting Coulomb coupling strength. The thermal noise could be more
critical in deteriorating the performance of the single-photon
router. Then, we also demonstrate the vacuum and thermal noise can
be insignificant for the optical performance of the single-photon
router at temperature of the order of 20 mK.

\emph{Model setup and the solutions}.----
\begin{figure}[tbp]
\centering
\includegraphics[width=8cm,height=4cm]{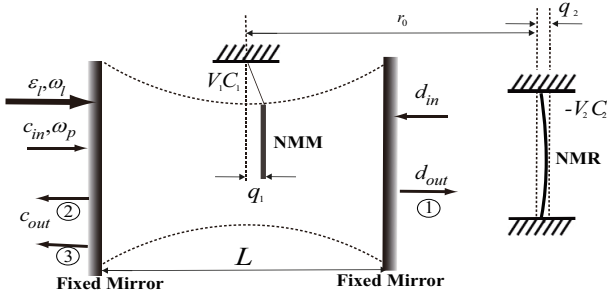}
\caption{Schematic diagram of the system. A high-quality
Fabry-Per\'{o}t cavity consists of two fixed mirrors and a NMM. NMM
is charged by the bias gate voltage $V_{1}$ and subject to the
Coulomb force due to the charged NMR  with the bias gate voltage
$V_{2}$. The optomechanical cavity of the length $L$ is driven by
two light fields, one of which is the pump field
$\protect\varepsilon_{l}$ with the frequency $\protect \omega_{l}$
and the other of which is the probe field $c_{in}$ with the
frequency $\protect\omega_{p}$. The output field is represented by
$c_{out}$ and $d_{out}$ . $q_{1}$ and $q_{2}$ represent the small
displacements of the NMM and NMR from their equilibrium positions,
with $r_{0}$ is the equilibrium distance between them.}
\end{figure}
The model for realizing single-photon routing is sketched in Fig. 1,
where a partially transparent nanomechanical mirror (NMM) which is
charged with the bias voltage $V_1$ placed at the middle position of
Fabry-Per\'{o}t cavity formed by two fixed mirrors that have finite
identical transmission \cite{nature-452-72}. The whole cavity length
is $L$. The cavity field is driven  by  a strong coupling field at
frequency $\omega_l$ from the left-hand mirror. Further, the field
in a single-photon Fock state at frequency $\omega_p$ is incident on
the cavity through the left-hand mirror. Besides the radiation
pressure force coupling the NMM to the cavity mode, the NMM is also
subject to Coulomb force to the charged NMR with the bias gate
voltage $-V_2$ near by the cavity. $q_1$ and $q_1$ represent the
small displacements of the NMM and the NMR around their equilibrium
positions, $r_0$ is the distance between their equilibrium position.
Then the Hamiltonian of the whole system can be written as
\begin{eqnarray}
H_{whole}&&=\hbar \omega _{c}c^{\dag
}c+\sum_{j=1}^2(\frac{p_{j}^{2}}{2m_{j}}+\frac{1}{2}
m_{j}\omega _{j}^{2}q_{j}^{2})  \notag \\
&&-\hbar gc^{\dag }cq_{1}+H_{I}+i\hbar \varepsilon _{l}(c^{\dag
}e^{-i\omega _{l}t}-H.C.),
\end{eqnarray}
where the first term is for the single-mode cavity field with
frequency $\omega _{c}$ and annihilation (creation) operator $c\
(c^{\dag })$. The second   term describes the vibration of the
charged NMM (NMR) with frequency $\omega_{1}$ ($\omega_{2}$) and
effective mass $m_{1}$ ($m_{2}$),  $p_{1}$ ($p_{2}$) and $q_{1}$
($q_{2}$) are  the momentum
 and the position operator of NMM (NMR), respectively \cite{pra-85-021801}.
The third term describe the radiation pressure coupling the cavity
field to the NMM, with $g=\frac{\omega_{c}}{L}$ is the coupling
strength \cite{nature-452-72}. The last   terms  describe the
interaction between the cavity field with the  input fields. The
pump field strength $\varepsilon_{l}$, depends on the power $\wp$ of
coupling field, $\varepsilon_{l}=\sqrt{2\kappa\wp_{l}/\omega_{l}}$
with $\kappa$ is the cavity decay rate.

The forth term $H_{I}$ in Eq.(1) presents the Coulomb coupling
between the charged NMM and NMR \cite{pra-72-041405}, where the NMM
and NMR take the charges $C_{1}V_{1}$ and $-C_{2}V_{2}$, with
$C_{1}$ and $C_{2}$   are the capacitance  of the   gates,
respectively. The interaction energy between the NMM and NMR is
given by $H_{I}=\frac{-C_{1}V_{1}C_{2}V_{2}}{4\pi \varepsilon
_{0}|r_{0}+q_{1}-q_{2}|}$. In the case of $q_{1},q_{2}\ll r_{0}$,
with the second order expansion, the above Hamiltonian is rewritten
as $H_{I}=\frac{-C_{1}V_{1}C_{2}V_{2}}{4\pi \varepsilon
_{0}r_{0}}[1-\frac{
q_{1}-q_{2}}{r_{0}}+(\frac{q_{1}-q_{2}}{r_{0}})^{2}]$, where the
linear term may be absorbed into the definition of the equilibrium
positions, and the quadratic term includes a renormalization of the
oscillation frequency for both the  NMM and NMR. It implies a
reduced form $H_{I}=\hbar \lambda q_{1}q_{2}$, where $\lambda
=\frac{C_{1}V_{1}C_{2}V_{2}}{2\pi \hbar \varepsilon _{0}r_{0}^{3}}$
\cite{pra-72-041405}.

In a frame rotating with the frequency $\omega_l$ of the pump field,
the Hamiltonian of the total system Eq.(1) can be rewritten as,
\begin{eqnarray}
H_{total}&&=\hbar\Delta_cc^\dag c
+\sum_{j=1}^2(\frac{p_{j}^{2}}{2m_{j}}+\frac{1}{2}
m_{j}\omega _{j}^{2}q_{j}^{2})  \notag \\
&&-\hbar gc^{\dag }cq_{1}+\hbar \lambda q_{1}q_{2}+i\hbar
\varepsilon _{l}(c^{\dag }-c),
\end{eqnarray}
where $\Delta_c=\omega_c-\omega_l$. Note that the NMM and the NR
coupled to the thermal surrounding at the temperature $T$, which
results in the mechanical damping rare $\gamma_1$ and $\gamma_2$,
and thermal noise force $\xi_1$ and $\xi_2$ with frequency-domain
correlation \cite{pra-85-021801},
\begin{eqnarray}
\langle \xi_\tau(\omega)\xi_\tau(\Omega)\rangle=2\pi\hbar\gamma_\tau
m_\tau \omega[1+
\coth(\frac{\hbar\omega}{2\kappa_BT})]\delta(\omega+\Omega),
\end{eqnarray}
where $\tau$=1 or 2 and $\kappa_B$ is the Boltzmann constant. In
addition, the cavity field $c$ is coupled to the input quantum
fields $c_{in}$ and $d_{in}$. If there are no photons incident from
the right, then $d_{in}$ would be the vacuum field. Let $2\kappa$ be
the rate at which photons leak out from each of the cavity mirrors.
The output  fields can be written as
\begin{eqnarray}
x_{out}(\omega)=\sqrt{2\kappa}c(\omega)-x_{in}(\omega),\ \  x=c\ , \
d.\label{4}
\end{eqnarray}
These couplings are included in the standard way by writing quantum
Langevin equations for the cavity field operators. Putting together
all the quantum fields, thermal fluctuations, and the Heisenberg
equations from the Hamiltonian (2), we can obtain the working
quantum Langevin equations:
\begin{eqnarray}
&&\dot{q_{1}}=\frac{p_{1}}{m_{1}},\ \ \ \  \dot{q_{2}}=\frac{p_{2}}{m_{2}}, \notag   \\
&&\dot{c}=-[2\kappa +i(\Delta _c-gq_1)]c+\varepsilon
_{l}+\sqrt{2\kappa}c_{in}+\sqrt{2\kappa}d_{in}, \notag \\
&&\dot{p_{1}}=-m_{1}\omega _{1}^{2}q_{1}-\hbar \lambda q_{2}+\hbar
gc^{\dag
}c-\gamma _{1}p_{1}+\xi _{1},  \notag \\
&&\dot{p_{2}}=-m_{2}\omega _{2}^{2}q_{2}-\hbar \lambda q_{1}-\gamma
_{2}p_{2}+\xi _{2},\label{5}
\end{eqnarray}
The quantum Langevin equations (\ref{5}) can be solved after all
operator are linearized as its steady-state mean value and a small
fluctuation:
\begin{eqnarray}
q_\tau=q_{\tau s}+\delta q_\tau,\  p_\tau=p_{\tau s}+\delta p_\tau,
\ c=c_s +\delta c, \label{6}
\end{eqnarray}
where $\delta q_\tau$, $\delta p_\tau$, $\delta c$ being the small
fluctuations around the corresponding steady values and $\tau$=1, 2.
After substituting Eq.(\ref{6}) into  Eq.(\ref{5}), ignoring the
second-order small terms, and introducing the Fourier transforms
$f(t)=\frac{1}{2\pi}\int_{-\infty}^{+\infty}f(\omega)e^{-i\omega
t}d\omega$,
$f^+(t)=\frac{1}{2\pi}\int_{-\infty}^{+\infty}f^+(-\omega)e^{-i\omega
t}d\omega$. We can get the steady values
\begin{eqnarray}
&&p_{1s}=p_{2s}=0, \ \   q_{1s}=\frac{\hbar
g|c_{s}|^{2}}{m_{1}\omega _{1}^{2}-\frac{\hbar
^{2}\lambda ^{2}}{m_{2}\omega _{2}^{2}}},  \notag \\
&&q_{2s}=\frac{\hbar \lambda q_{1s}}{-m_{2}\omega _{2}^{2}}, \ \
c_{s}=\frac{\varepsilon _{l}}{2\kappa+ i\Delta },
\end{eqnarray}
with $\Delta =\Delta _{c}-gq_{1s}$, and the solution of $\delta c$
\cite{pra-83-043826},
\begin{eqnarray}
\delta c
=E(\omega)c_{in}(\omega)+F(\omega)c^+_{in}(-\omega)+E(\omega)d_{in}(\omega)
\nonumber \\
+F(\omega)d^+_{in}(-\omega)+V_1(\omega)\xi_1(\omega)+V_2(\omega)\xi_2(\omega),\label{8}
\end{eqnarray}
in which
\begin{eqnarray}
&&E(\omega)=\sqrt{2\kappa}[\frac{1}{2\kappa+i(\Delta-\omega)}
\nonumber \\
&&\ \ \ \ \ \ \ \ +\frac{ig^2\hbar|c_s|^2(2\kappa
i+\Delta+\omega)m_2(\omega^2+i\omega\gamma_2-\omega_2^2)}{d(\omega)}],
\nonumber \\
&&F(\omega)=\frac{i\sqrt{2\kappa}g^2\hbar|c_s|^2(2\kappa
i-\Delta+\omega)m_2(\omega^2+i\omega\gamma_2-\omega_2^2)}{d(\omega)},\nonumber \\
&&V_1(\omega)=\frac{g|c_s|^2[(2\kappa i+\omega)^2-\Delta^2]m_2(\omega^2+i\omega\gamma_2-\omega_2^2)}{d(\omega)},\nonumber \\
&&V_2(\omega)=\frac{g\hbar\lambda|c_s|^2[(2\kappa
i+\omega)^2-\Delta^2]}{d(\omega)},
\end{eqnarray}
with
\begin{eqnarray}
&&d(\omega)=(2\kappa i+
\Delta-\omega)\{-\hbar^2\lambda^2[\Delta^2+(2\kappa
i+\omega)^2]\nonumber \\
&&+m_2(\omega^2+i\omega\gamma_2-\omega_2^2)[2|c_s|^2g^2\hbar\Delta+(\Delta-2\kappa
i-\omega)\nonumber \\
&&\times(\Delta+2\kappa
i+\omega)m_1(\omega^2+i\omega\gamma_1-\omega_1^2)] \}.
\end{eqnarray}

\begin{figure}
\begin{minipage}[b]{0.5 \textwidth}
\includegraphics[width=0.5 \textwidth]{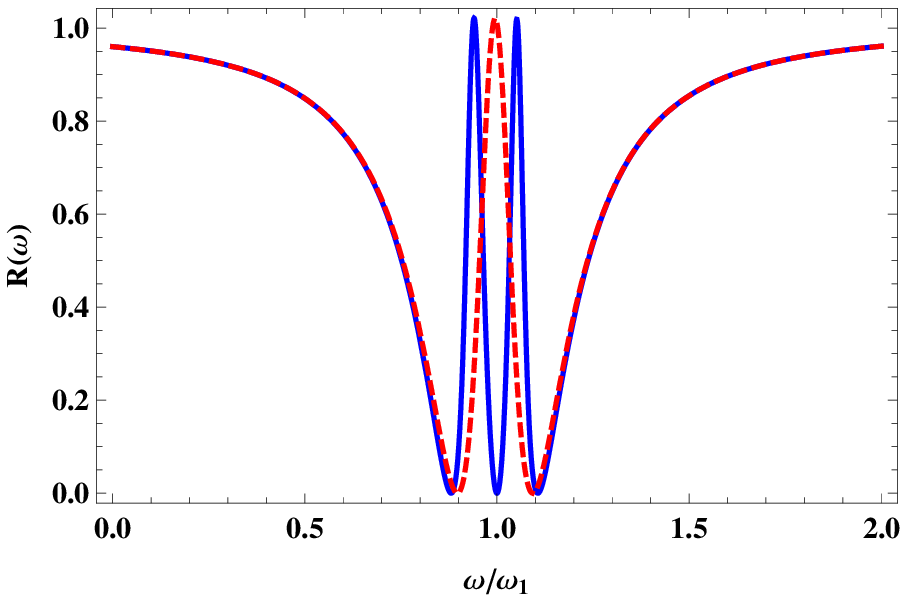}\includegraphics[width=0.5 \textwidth]{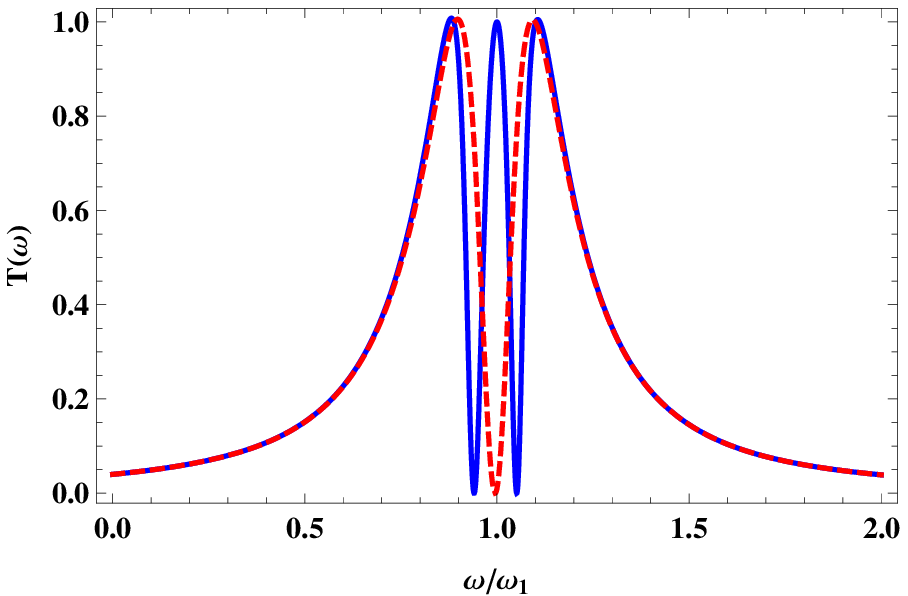}
\end{minipage}
\caption{(Color online) (a)  The reflection spectrum $R(\omega)$ and
(b) the transmission spectrum $T(\omega)$  of the single photon as a
function of normalized frequency $\omega/\omega_1$ when the Coulomb
interaction is turned off (red dashed line) or turned on (blue solid
line).$\protect\lambda_l=$1054 nm, $L=$6.7 cm,
$\omega_1=\omega_2=2\pi\times134\times10^3$Hz,
$Q_1=Q_2=1.1\times10^6$, $m_1=m_2=40$ ng,
$\protect\kappa=\omega_1/10$, $\wp_l=2\ \mu$W, and
$\protect\lambda=3\times10^{33}$ Hz/m$^2$. }
\end{figure}

Defining the spectrum  of the field via $\langle
c^+(-\Omega)c(\omega)\rangle=2\pi S_c(\omega)\delta(\omega+\Omega)$,
$\langle c(\omega)c^+(-\Omega)\rangle=2\pi
[S_c(\omega)+1]\delta(\omega+\Omega)$. The incoming vacuum field
$d_{in}$ is characterized by $\langle
d(\omega)d^+(-\Omega)\rangle=2\pi\delta(\omega+\Omega)$ with
$S_{din}(\omega)=0$. From Eq. (\ref{4}) and Eq. (\ref{8}), we can
obtain the spectrum of the output  fields,
\begin{eqnarray}
S_{cout}(\omega)=R(\omega)S_{cin}+S^{(v)}(\omega)+S_1^{T}(\omega)+S_2^{T}(\omega), \nonumber \\
S_{dout}(\omega)=T(\omega)S_{cin}+S^{(v)}(\omega)+S_1^{T}(\omega)+S_2^{T}(\omega),\label{11}
\end{eqnarray}
where
\begin{eqnarray}
&&R(\omega)=|\sqrt{2\kappa}E(\omega)-1|^2, \ \ \ \
T(\omega)=|\sqrt{2\kappa}E(\omega)|^2 ,\nonumber \\
&&S_1^{T}(\omega)=2\kappa |V_1(\omega)|^2 \hbar\gamma_1
m_1(-\omega)[1+
\coth(\frac{-\hbar\omega}{2\kappa_BT})] \nonumber \\
&&S_2^{T}(\omega)=2\kappa |V_2(\omega)|^2 \hbar\gamma_2
m_2(-\omega)[1+ \coth(\frac{-\hbar\omega}{2\kappa_BT})],\nonumber \\
&&S^{(v)}(\omega)=4\kappa |F(\omega)|^2.
\end{eqnarray}
When there is no Coulomb coupling $\lambda$ (\emph{i.e.}
$\lambda=0$) between the NMM and the NMR, Eq.(\ref{11}) can be
reduced to Eq.(10) in Ref.\cite{pra-85-021801}. However, different
from the output field in Ref.\cite{pra-85-021801} involving a single
output channel for quantum router, there are three output channels
with different frequencies in our scheme due to Coulomb interaction.
Further more, the frequencies of two output channel can be selected
by adjusting the Coulomb coupling  strength $\lambda$.

\emph{Quantum routing for single photons}. ----  In Eq.(\ref{11}),
$R(\omega)$ and  $T(\omega)$ are the contributions arising from the
presence of a single photon in the input field. $S^{(v)}(\omega)$ is
the contribution from incoming vacuum field. The $S_1^{T}(\omega)$
and $S_2^{T}(\omega)$ are contributions from the fluctuation of the
NMM and the NMR, respectively. Eq.(\ref{11}) shows that even if
there were no incoming photon, the output signal is generated via
quantum and thermal noises. For the purpose of achieving a
single-photon router, the key quantities are $R(\omega)$ and
$T(\omega)$. Further, we also demonstrate the performance of the
single-photon router should not be deteriorated by the quantum and
thermal noises terms $S^{(v)}(\omega)$ and $S_\tau^{T}(\omega)\
(\tau=1,\  2)$.

To demonstrate the routing functions of our hybrid optomechanics
system, we first investigate the reflection $R(\omega)$ and
transmission spectrum $T(\omega)$. For illustration of the numerical
results, we choose the realistically reasonable parameters from the
recent experiment \cite{nature-452-72} on the observation of the
strong dispersive coupling between a cavity and the NMM. The
wavelength of the pump field $\lambda_l=1054 \ $nm, $L=6.7\ $cm,
$m_1=m_2=40 \ $ng, $\omega_1=\omega_2=2\pi\times134\times10^3\ $Hz,
quality factor $Q_{1}=Q_{2}=1.1\times10^6$,
$\gamma_1=\gamma_2=\frac{\omega_1}{Q_1}=0.76\ $s$^{-1}$,
$\kappa=\frac{\omega_1}{10}$, $\wp_l=5\ \mu$W, we choose coulomb
coupling strength $\lambda=3\times10^{33}$Hz/m$^2$. We also apply
the following conditions \cite{pra-81-041803, science-330-1520}, (i)
$\Delta\simeq\omega_1$ and (ii) $\omega_1\gg\kappa$. The first
condition means that the optical cavity is driven by a red-detuned
laser field which is on resonance with the optomechanical
anti-Stokes sideband. The second condition is the well-known
resolved sideband condition, which ensures the normal mode splitting
to be distinguished \cite{science-330-1520}.

\begin{figure}[tbp]
\centering
\includegraphics[width=7cm,height=4cm]{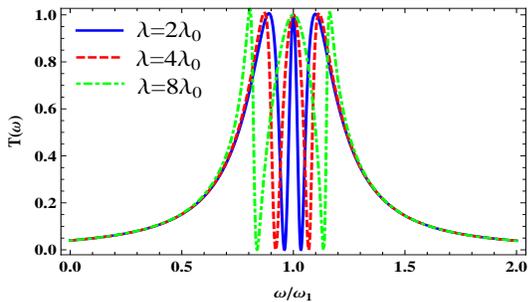}
\caption{(Color online) The transmission spectrum $T(\omega)$ of the
single photon as a function of normalized frequency
$\omega/\omega_1$  with different $\lambda$ (in units of
$\lambda_0$, $\lambda_0=10^{33}$ Hz/m$^2$). Other parameters take
the same values as in Fig. 2.}
\end{figure}

The resulting spectra are shown in Figs 2 and 3. In the absence of
the Coulomb coupling strength (\emph{i.e.} $\lambda=0$), one
observes an inverted EIT and a standard EIT in the reflection and
transmission spectra of a single photon. Note the
$R(\omega_1)\approx1$ and $T(\omega)\approx0$. So the single photon
is complete reflected through the cavity to the left output port.
However, in the presence of the Coulomb coupling strength, the
situation is completely different $R(\omega_1)\approx0$ and
$T(\omega)\approx1$. More important, the reflection and transmission
spectra of a single photon exhibit other two inverted dips and two
normal dips at $\omega=\omega_1+\omega_0$ and
$\omega=\omega_1-\omega_0$, here $\omega_0$ is the small deviation
from the central frequency $\omega_1$ and it  depend on the Coulomb
coupling strength. We can find $R(\omega_1\pm\omega_0)\approx1$ and
$T(\omega_1\pm\omega_0)\approx0$. That is to say, the single photon
is completely transmitted through the cavity to the right output
port at frequency $\omega=\omega_1$, meanwhile, the single photon is
completely reflected  to the left two output ports with different
frequencies $\omega_1+\omega_0$ and $\omega_1-\omega_0$.

Then, we can describe the working process of the single-photon
multi-output router. When we turn off the Coulomb coupling,   the
single photons is complete reflected through the cavity to the left
output port at frequency $\omega=\omega_1$ (\emph{i.e.}
$R(\omega_1)\approx1$, $T(\omega)\approx0$). However, When we turn
on the Coulomb coupling, the single photons is complete transmitted
to the right output port at frequency $\omega=\omega_1$ (
\emph{i.e.} $R(\omega_1)\approx0$, $T(\omega)\approx1$), at the same
time, three are completely reflected to the left two outputs at
different frequencies $\omega=\omega_1+\omega_0$ and
$\omega=\omega_1-\omega_0$ (\emph{i.e.}
$R(\omega_1\pm\omega_0)\approx1$ and
$T(\omega_1\pm\omega_0)\approx0$). Fig. 4 describe the transmission
spectrum $T(\omega)$ with different Coulomb coupling strength. From
Fig. 4, we find the different output frequencies can be selected by
adjusting the Coulomb coupling strength.

\begin{figure}
\begin{minipage}[b]{0.5 \textwidth}
\includegraphics[width=0.5 \textwidth]{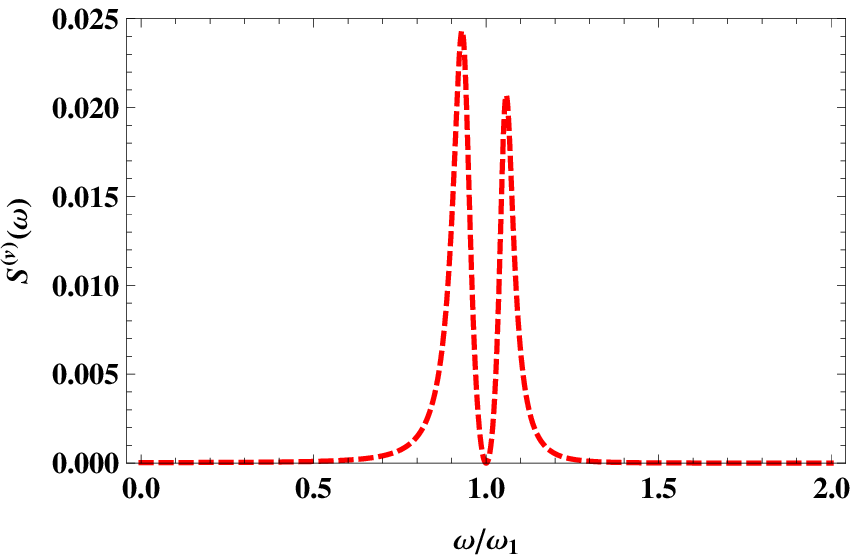}\includegraphics[width=0.5 \textwidth]{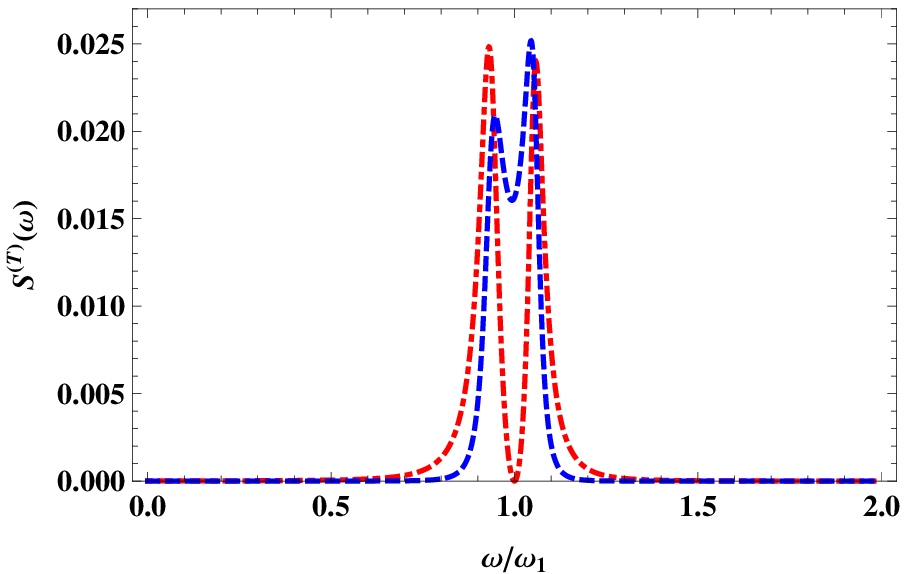}
\end{minipage}
\caption{ (a) The vacuum noise spectrum $S^{(v)}(\omega)$ as a
function of normalized frequency $\omega/\omega_1$.  (b) The thermal
noise spectrum $S_1^{(T)}(\omega)$ (red dotdashed line) and
$S_2^{(T)}(\omega)$ (blue dashed line)   as a function of normalized
frequency $\omega/\omega_1$.  $T=20$mK, other parameters take the
same values as in Fig. 2.}
\end{figure}

Next, we discuss the effects of the quantum and thermal noise on the
reflection and transmission spectrum of a single-photon. From Fig.5,
the contribution of the vacuum noise maximum is about $2.5\%$  and
is thus insignificant. The thermal noise could be more critical in
deteriorating the performance of the single-photon router. Clearly
to beat the effects of thermal noise, the number of photons in the
probe pulse has to be much bigger than the thermal noise photons.
However, if we work with NMM and NMR temperatures like 20 mK, then
the thermal noise term is insignificant as shown in Fig. 6.

\emph{In conclusion.}----We have proposed an experimentally
accessible tunable single-photon multi-channel routing scheme based
on the hybrid optoelectromechanical system which consist of an
high-quality Fabry-Per\'{o}t cavity  coulomb coupling to a NMR. Our
proposal is essentially different from previous methods in
Ref.\cite{prx-3-031013,prl-111-103604,pra-89-013806} that we are
able to directly modulate the coherent input single-photon signal
into three different output ports with unit probability. More
important, the two output signal frequencies can be selected by
adjusting Coulomb coupling strength. We also demonstrate the vacuum
and thermal noise will be insignificant for the optical performance
of the single-photon router at temperature of the order of 20 mK. We
hope that the quantum routing function prodicted in this letter can
be observed in some experiments. Our proposal may have paved a new
avenue towards multi-channel router and quantum networks.

\section*{ACKNOWLEDGMENTS}
This work was supported by the Major Research Plan of the NSFC
(Grant No.91121023), the NSFC (Grants No. 61378012, No. 60978009,
No. 11274352 and No. 11304366), the SRFDPHEC(Grant
No.20124407110009), the "973"Program (Grant Nos. 2011CBA00200,
2012CB922102 and 2013CB921804), the PCSIRT (Grant No.IRT1243). China
Postdoctoral Science Foundation (Grant No. 2013M531771).  Natural
Science Fund for colleges and universities in Jiangsu Province
(Grant No.12KJD140002).

\end{document}